\hsize=31pc
\vsize=49pc
\lineskip=0pt
\parskip=0pt plus 1pt
\hfuzz=1pt
\vfuzz=2pt
\pretolerance=2500
\tolerance=5000
\vbadness=5000
\hbadness=5000
\widowpenalty=500
\clubpenalty=200
\brokenpenalty=500
\predisplaypenalty=200
\voffset=-1pc
\nopagenumbers
\catcode`@=11
\newif\ifams
\amstrue
%
%
\global\amsfalse
%
%
\newfam\bdifam
\newfam\bsyfam
\newfam\bssfam
\newfam\msafam
\newfam\msbfam
\newif\ifxxpt
\newif\ifxviipt
\newif\ifxivpt
\newif\ifxiipt
\newif\ifxipt
\newif\ifxpt
\newif\ifixpt
\newif\ifviiipt
\newif\ifviipt
\newif\ifvipt
\newif\ifvpt
%
%
\def\headsize#1#2{\def\headb@seline{#2}%
                \ifnum#1=20\def\HEAD{twenty}%
                           \def\smHEAD{twelve}%
                           \def\vsHEAD{nine}%
                           \ifxxpt\else\xdef\f@ntsize{\HEAD}%
                           \def\m@g{4}\def\s@ze{20.74}%
                           \loadheadfonts\xxpttrue\fi
                           \ifxiipt\else\xdef\f@ntsize{\smHEAD}%
                           \def\m@g{1}\def\s@ze{12}%
                           \loadxiiptfonts\xiipttrue\fi
                           \ifixpt\else\xdef\f@ntsize{\vsHEAD}%
                           \def\s@ze{9}%
                           \loadsmallfonts\ixpttrue\fi
                      \else
                \ifnum#1=17\def\HEAD{seventeen}%
                           \def\smHEAD{eleven}%
                           \def\vsHEAD{eight}%
                           \ifxviipt\else\xdef\f@ntsize{\HEAD}%
                           \def\m@g{3}\def\s@ze{17.28}%
                           \loadheadfonts\xviipttrue\fi
                           \ifxipt\else\xdef\f@ntsize{\smHEAD}%
                           \loadxiptfonts\xipttrue\fi
                           \ifviiipt\else\xdef\f@ntsize{\vsHEAD}%
                           \def\s@ze{8}%
                           \loadsmallfonts\viiipttrue\fi
                      \else\def\HEAD{fourteen}%
                           \def\smHEAD{ten}%
                           \def\vsHEAD{seven}%
                           \ifxivpt\else\xdef\f@ntsize{\HEAD}%
                           \def\m@g{2}\def\s@ze{14.4}%
                           \loadheadfonts\xivpttrue\fi
                           \ifxpt\else\xdef\f@ntsize{\smHEAD}%
                           \def\s@ze{10}%
                           \loadxptfonts\xpttrue\fi
                           \ifviipt\else\xdef\f@ntsize{\vsHEAD}%
                           \def\s@ze{7}%
                           \loadviiptfonts\viipttrue\fi
                \ifnum#1=14\else
                \message{Header size should be 20, 17 or 14 point
                              will now default to 14pt}\fi
                \fi\fi\headfonts}
%
%
\def\textsize#1#2{\def\textb@seline{#2}%
                 \ifnum#1=12\def\TEXT{twelve}%
                           \def\smTEXT{eight}%
                           \def\vsTEXT{six}%
                           \ifxiipt\else\xdef\f@ntsize{\TEXT}%
                           \def\m@g{1}\def\s@ze{12}%
                           \loadxiiptfonts\xiipttrue\fi
                           \ifviiipt\else\xdef\f@ntsize{\smTEXT}%
                           \def\s@ze{8}%
                           \loadsmallfonts\viiipttrue\fi
                           \ifvipt\else\xdef\f@ntsize{\vsTEXT}%
                           \def\s@ze{6}%
                           \loadviptfonts\vipttrue\fi
                      \else
                \ifnum#1=11\def\TEXT{eleven}%
                           \def\smTEXT{seven}%
                           \def\vsTEXT{five}%
                           \ifxipt\else\xdef\f@ntsize{\TEXT}%
                           \def\s@ze{11}%
                           \loadxiptfonts\xipttrue\fi
                           \ifviipt\else\xdef\f@ntsize{\smTEXT}%
                           \loadviiptfonts\viipttrue\fi
                           \ifvpt\else\xdef\f@ntsize{\vsTEXT}%
                           \def\s@ze{5}%
                           \loadvptfonts\vpttrue\fi
                      \else\def\TEXT{ten}%
                           \def\smTEXT{seven}%
                           \def\vsTEXT{five}%
                           \ifxpt\else\xdef\f@ntsize{\TEXT}%
                           \loadxptfonts\xpttrue\fi
                           \ifviipt\else\xdef\f@ntsize{\smTEXT}%
                           \def\s@ze{7}%
                           \loadviiptfonts\viipttrue\fi
                           \ifvpt\else\xdef\f@ntsize{\vsTEXT}%
                           \def\s@ze{5}%
                           \loadvptfonts\vpttrue\fi
                \ifnum#1=10\else
                \message{Text size should be 12, 11 or 10 point
                              will now default to 10pt}\fi
                \fi\fi\textfonts}
%
%
\def\smallsize#1#2{\def\smallb@seline{#2}%
                 \ifnum#1=10\def\SMALL{ten}%
                           \def\smSMALL{seven}%
                           \def\vsSMALL{five}%
                           \ifxpt\else\xdef\f@ntsize{\SMALL}%
                           \loadxptfonts\xpttrue\fi
                           \ifviipt\else\xdef\f@ntsize{\smSMALL}%
                           \def\s@ze{7}%
                           \loadviiptfonts\viipttrue\fi
                           \ifvpt\else\xdef\f@ntsize{\vsSMALL}%
                           \def\s@ze{5}%
                           \loadvptfonts\vpttrue\fi
                       \else
                 \ifnum#1=9\def\SMALL{nine}%
                           \def\smSMALL{six}%
                           \def\vsSMALL{five}%
                           \ifixpt\else\xdef\f@ntsize{\SMALL}%
                           \def\s@ze{9}%
                           \loadsmallfonts\ixpttrue\fi
                           \ifvipt\else\xdef\f@ntsize{\smSMALL}%
                           \def\s@ze{6}%
                           \loadviptfonts\vipttrue\fi
                           \ifvpt\else\xdef\f@ntsize{\vsSMALL}%
                           \def\s@ze{5}%
                           \loadvptfonts\vpttrue\fi
                       \else
                           \def\SMALL{eight}%
                           \def\smSMALL{six}%
                           \def\vsSMALL{five}%
                           \ifviiipt\else\xdef\f@ntsize{\SMALL}%
                           \def\s@ze{8}%
                           \loadsmallfonts\viiipttrue\fi
                           \ifvipt\else\xdef\f@ntsize{\smSMALL}%
                           \def\s@ze{6}%
                           \loadviptfonts\vipttrue\fi
                           \ifvpt\else\xdef\f@ntsize{\vsSMALL}%
                           \def\s@ze{5}%
                           \loadvptfonts\vpttrue\fi
                 \ifnum#1=8\else\message{Small size should be 10, 9 or
                            8 point will now default to 8pt}\fi
                \fi\fi\smallfonts}
\def\F@nt{\expandafter\font\csname}
\def\Sk@w{\expandafter\skewchar\csname}
\def\@nd{\endcsname}
\def\@step#1{ scaled \magstep#1}
\def\@half{ scaled \magstephalf}
\def\@t#1{ at #1pt}
%
%
\def\loadheadfonts{\bigf@nts
\F@nt \f@ntsize bdi\@nd=cmmib10 \@t{\s@ze}%
\Sk@w \f@ntsize bdi\@nd='177
\F@nt \f@ntsize bsy\@nd=cmbsy10 \@t{\s@ze}%
\Sk@w \f@ntsize bsy\@nd='60
\F@nt \f@ntsize bss\@nd=cmssbx10 \@t{\s@ze}}
%
%
\def\loadxiiptfonts{\bigf@nts
\F@nt \f@ntsize bdi\@nd=cmmib10 \@step{\m@g}%
\Sk@w \f@ntsize bdi\@nd='177
\F@nt \f@ntsize bsy\@nd=cmbsy10 \@step{\m@g}%
\Sk@w \f@ntsize bsy\@nd='60
\F@nt \f@ntsize bss\@nd=cmssbx10 \@step{\m@g}}
%
%
\def\loadxiptfonts{%
\font\elevenrm=cmr10 \@half
\font\eleveni=cmmi10 \@half
\skewchar\eleveni='177
\font\elevensy=cmsy10 \@half
\skewchar\elevensy='60
\font\elevenex=cmex10 \@half
\font\elevenit=cmti10 \@half
\font\elevensl=cmsl10 \@half
\font\elevenbf=cmbx10 \@half
\font\eleventt=cmtt10 \@half
\ifams\font\elevenmsa=msam10 \@half
\font\elevenmsb=msbm10 \@half\else\fi
\font\elevenbdi=cmmib10 \@half
\skewchar\elevenbdi='177
\font\elevenbsy=cmbsy10 \@half
\skewchar\elevenbsy='60
\font\elevenbss=cmssbx10 \@half}
%
%
\def\loadxptfonts{%
\font\tenbdi=cmmib10
\skewchar\tenbdi='177
\font\tenbsy=cmbsy10
\skewchar\tenbsy='60
\ifams\font\tenmsa=msam10
\font\tenmsb=msbm10\else\fi
\font\tenbss=cmssbx10}%
%
%
\def\loadsmallfonts{\smallf@nts
\ifams
\F@nt \f@ntsize ex\@nd=cmex\s@ze
\else
\F@nt \f@ntsize ex\@nd=cmex10\fi
\F@nt \f@ntsize it\@nd=cmti\s@ze
\F@nt \f@ntsize sl\@nd=cmsl\s@ze
\F@nt \f@ntsize tt\@nd=cmtt\s@ze}
%
%
\def\loadviiptfonts{%
\font\sevenit=cmti7
\font\sevensl=cmsl8 at 7pt
\ifams\font\sevenmsa=msam7
\font\sevenmsb=msbm7
\font\sevenex=cmex7
\font\sevenbsy=cmbsy7
\font\sevenbdi=cmmib7\else
\font\sevenex=cmex10
\font\sevenbsy=cmbsy10 at 7pt
\font\sevenbdi=cmmib10 at 7pt\fi
\skewchar\sevenbsy='60
\skewchar\sevenbdi='177
\font\sevenbss=cmssbx10 at 7pt}%
%
%
\def\loadviptfonts{\smallf@nts
\ifams\font\sixex=cmex7 at 6pt\else
\font\sixex=cmex10\fi
\font\sixit=cmti7 at 6pt}
%
%
\def\loadvptfonts{%
\font\fiveit=cmti7 at 5pt
\ifams\font\fiveex=cmex7 at 5pt
\font\fivebdi=cmmib5
\font\fivebsy=cmbsy5
\font\fivemsa=msam5
\font\fivemsb=msbm5\else
\font\fiveex=cmex10
\font\fivebdi=cmmib10 at 5pt
\font\fivebsy=cmbsy10 at 5pt\fi
\skewchar\fivebdi='177
\skewchar\fivebsy='60
\font\fivebss=cmssbx10 at 5pt}
\def\bigf@nts{%
\F@nt \f@ntsize rm\@nd=cmr10 \@step{\m@g}%
\F@nt \f@ntsize i\@nd=cmmi10 \@step{\m@g}%
\Sk@w \f@ntsize i\@nd='177
\F@nt \f@ntsize sy\@nd=cmsy10 \@step{\m@g}%
\Sk@w \f@ntsize sy\@nd='60
\F@nt \f@ntsize ex\@nd=cmex10 \@step{\m@g}%
\F@nt \f@ntsize it\@nd=cmti10 \@step{\m@g}%
\F@nt \f@ntsize sl\@nd=cmsl10 \@step{\m@g}%
\F@nt \f@ntsize bf\@nd=cmbx10 \@step{\m@g}%
\F@nt \f@ntsize tt\@nd=cmtt10 \@step{\m@g}%
\ifams
\F@nt \f@ntsize msa\@nd=msam10 \@step{\m@g}%
\F@nt \f@ntsize msb\@nd=msbm10 \@step{\m@g}\else\fi}
\def\smallf@nts{%
\F@nt \f@ntsize rm\@nd=cmr\s@ze
\F@nt \f@ntsize i\@nd=cmmi\s@ze
\Sk@w \f@ntsize i\@nd='177
\F@nt \f@ntsize sy\@nd=cmsy\s@ze
\Sk@w \f@ntsize sy\@nd='60
\F@nt \f@ntsize bf\@nd=cmbx\s@ze
\ifams
\F@nt \f@ntsize bdi\@nd=cmmib\s@ze
\F@nt \f@ntsize bsy\@nd=cmbsy\s@ze
\F@nt \f@ntsize msa\@nd=msam\s@ze
\F@nt \f@ntsize msb\@nd=msbm\s@ze
\else
\F@nt \f@ntsize bdi\@nd=cmmib10 \@t{\s@ze}%
\F@nt \f@ntsize bsy\@nd=cmbsy10 \@t{\s@ze}\fi
\Sk@w \f@ntsize bdi\@nd='177
\Sk@w \f@ntsize bsy\@nd='60
\F@nt \f@ntsize bss\@nd=cmssbx10 \@t{\s@ze}}%
%
%
\def\headfonts{%
\textfont0=\csname\HEAD rm\@nd
\scriptfont0=\csname\smHEAD rm\@nd
\scriptscriptfont0=\csname\vsHEAD rm\@nd
\def\rm{\fam0\csname\HEAD rm\@nd
\def\sc{\csname\smHEAD rm\@nd}}%
\textfont1=\csname\HEAD i\@nd
\scriptfont1=\csname\smHEAD i\@nd
\scriptscriptfont1=\csname\vsHEAD i\@nd
\textfont2=\csname\HEAD sy\@nd
\scriptfont2=\csname\smHEAD sy\@nd
\scriptscriptfont2=\csname\vsHEAD sy\@nd
\textfont3=\csname\HEAD ex\@nd
\scriptfont3=\csname\smHEAD ex\@nd
\scriptscriptfont3=\csname\smHEAD ex\@nd
\textfont\itfam=\csname\HEAD it\@nd
\scriptfont\itfam=\csname\smHEAD it\@nd
\scriptscriptfont\itfam=\csname\vsHEAD it\@nd
\def\it{\fam\itfam\csname\HEAD it\@nd
\def\sc{\csname\smHEAD it\@nd}}%
\textfont\slfam=\csname\HEAD sl\@nd
\def\sl{\fam\slfam\csname\HEAD sl\@nd
\def\sc{\csname\smHEAD sl\@nd}}%
\textfont\bffam=\csname\HEAD bf\@nd
\scriptfont\bffam=\csname\smHEAD bf\@nd
\scriptscriptfont\bffam=\csname\vsHEAD bf\@nd
\def\bf{\fam\bffam\csname\HEAD bf\@nd
\def\sc{\csname\smHEAD bf\@nd}}%
\textfont\ttfam=\csname\HEAD tt\@nd
\def\tt{\fam\ttfam\csname\HEAD tt\@nd}%
\textfont\bdifam=\csname\HEAD bdi\@nd
\scriptfont\bdifam=\csname\smHEAD bdi\@nd
\scriptscriptfont\bdifam=\csname\vsHEAD bdi\@nd
\def\bdi{\fam\bdifam\csname\HEAD bdi\@nd}%
\textfont\bsyfam=\csname\HEAD bsy\@nd
\scriptfont\bsyfam=\csname\smHEAD bsy\@nd
\def\bsy{\fam\bsyfam\csname\HEAD bsy\@nd}%
\textfont\bssfam=\csname\HEAD bss\@nd
\scriptfont\bssfam=\csname\smHEAD bss\@nd
\scriptscriptfont\bssfam=\csname\vsHEAD bss\@nd
\def\bss{\fam\bssfam\csname\HEAD bss\@nd}%
\ifams
\textfont\msafam=\csname\HEAD msa\@nd
\scriptfont\msafam=\csname\smHEAD msa\@nd
\scriptscriptfont\msafam=\csname\vsHEAD msa\@nd
\textfont\msbfam=\csname\HEAD msb\@nd
\scriptfont\msbfam=\csname\smHEAD msb\@nd
\scriptscriptfont\msbfam=\csname\vsHEAD msb\@nd
\else\fi
\normalbaselineskip=\headb@seline pt%
\setbox\strutbox=\hbox{\vrule height.7\normalbaselineskip
depth.3\baselineskip width0pt}%
\def\sc{\csname\smHEAD rm\@nd}\normalbaselines\bf}
%
%
\def\textfonts{%
\textfont0=\csname\TEXT rm\@nd
\scriptfont0=\csname\smTEXT rm\@nd
\scriptscriptfont0=\csname\vsTEXT rm\@nd
\def\rm{\fam0\csname\TEXT rm\@nd
\def\sc{\csname\smTEXT rm\@nd}}%
\textfont1=\csname\TEXT i\@nd
\scriptfont1=\csname\smTEXT i\@nd
\scriptscriptfont1=\csname\vsTEXT i\@nd
\textfont2=\csname\TEXT sy\@nd
\scriptfont2=\csname\smTEXT sy\@nd
\scriptscriptfont2=\csname\vsTEXT sy\@nd
\textfont3=\csname\TEXT ex\@nd
\scriptfont3=\csname\smTEXT ex\@nd
\scriptscriptfont3=\csname\smTEXT ex\@nd
\textfont\itfam=\csname\TEXT it\@nd
\scriptfont\itfam=\csname\smTEXT it\@nd
\scriptscriptfont\itfam=\csname\vsTEXT it\@nd
\def\it{\fam\itfam\csname\TEXT it\@nd
\def\sc{\csname\smTEXT it\@nd}}%
\textfont\slfam=\csname\TEXT sl\@nd
\def\sl{\fam\slfam\csname\TEXT sl\@nd
\def\sc{\csname\smTEXT sl\@nd}}%
\textfont\bffam=\csname\TEXT bf\@nd
\scriptfont\bffam=\csname\smTEXT bf\@nd
\scriptscriptfont\bffam=\csname\vsTEXT bf\@nd
\def\bf{\fam\bffam\csname\TEXT bf\@nd
\def\sc{\csname\smTEXT bf\@nd}}%
\textfont\ttfam=\csname\TEXT tt\@nd
\def\tt{\fam\ttfam\csname\TEXT tt\@nd}%
\textfont\bdifam=\csname\TEXT bdi\@nd
\scriptfont\bdifam=\csname\smTEXT bdi\@nd
\scriptscriptfont\bdifam=\csname\vsTEXT bdi\@nd
\def\bdi{\fam\bdifam\csname\TEXT bdi\@nd}%
\textfont\bsyfam=\csname\TEXT bsy\@nd
\scriptfont\bsyfam=\csname\smTEXT bsy\@nd
\def\bsy{\fam\bsyfam\csname\TEXT bsy\@nd}%
\textfont\bssfam=\csname\TEXT bss\@nd
\scriptfont\bssfam=\csname\smTEXT bss\@nd
\scriptscriptfont\bssfam=\csname\vsTEXT bss\@nd
\def\bss{\fam\bssfam\csname\TEXT bss\@nd}%
\ifams
\textfont\msafam=\csname\TEXT msa\@nd
\scriptfont\msafam=\csname\smTEXT msa\@nd
\scriptscriptfont\msafam=\csname\vsTEXT msa\@nd
\textfont\msbfam=\csname\TEXT msb\@nd
\scriptfont\msbfam=\csname\smTEXT msb\@nd
\scriptscriptfont\msbfam=\csname\vsTEXT msb\@nd
\else\fi
\normalbaselineskip=\textb@seline pt
\setbox\strutbox=\hbox{\vrule height.7\normalbaselineskip
depth.3\baselineskip width0pt}%
\everymath{}%
\def\sc{\csname\smTEXT rm\@nd}\normalbaselines\rm}
%
%
\def\smallfonts{%
\textfont0=\csname\SMALL rm\@nd
\scriptfont0=\csname\smSMALL rm\@nd
\scriptscriptfont0=\csname\vsSMALL rm\@nd
\def\rm{\fam0\csname\SMALL rm\@nd
\def\sc{\csname\smSMALL rm\@nd}}%
\textfont1=\csname\SMALL i\@nd
\scriptfont1=\csname\smSMALL i\@nd
\scriptscriptfont1=\csname\vsSMALL i\@nd
\textfont2=\csname\SMALL sy\@nd
\scriptfont2=\csname\smSMALL sy\@nd
\scriptscriptfont2=\csname\vsSMALL sy\@nd
\textfont3=\csname\SMALL ex\@nd
\scriptfont3=\csname\smSMALL ex\@nd
\scriptscriptfont3=\csname\smSMALL ex\@nd
\textfont\itfam=\csname\SMALL it\@nd
\scriptfont\itfam=\csname\smSMALL it\@nd
\scriptscriptfont\itfam=\csname\vsSMALL it\@nd
\def\it{\fam\itfam\csname\SMALL it\@nd
\def\sc{\csname\smSMALL it\@nd}}%
\textfont\slfam=\csname\SMALL sl\@nd
\def\sl{\fam\slfam\csname\SMALL sl\@nd
\def\sc{\csname\smSMALL sl\@nd}}%
\textfont\bffam=\csname\SMALL bf\@nd
\scriptfont\bffam=\csname\smSMALL bf\@nd
\scriptscriptfont\bffam=\csname\vsSMALL bf\@nd
\def\bf{\fam\bffam\csname\SMALL bf\@nd
\def\sc{\csname\smSMALL bf\@nd}}%
\textfont\ttfam=\csname\SMALL tt\@nd
\def\tt{\fam\ttfam\csname\SMALL tt\@nd}%
\textfont\bdifam=\csname\SMALL bdi\@nd
\scriptfont\bdifam=\csname\smSMALL bdi\@nd
\scriptscriptfont\bdifam=\csname\vsSMALL bdi\@nd
\def\bdi{\fam\bdifam\csname\SMALL bdi\@nd}%
\textfont\bsyfam=\csname\SMALL bsy\@nd
\scriptfont\bsyfam=\csname\smSMALL bsy\@nd
\def\bsy{\fam\bsyfam\csname\SMALL bsy\@nd}%
\textfont\bssfam=\csname\SMALL bss\@nd
\scriptfont\bssfam=\csname\smSMALL bss\@nd
\scriptscriptfont\bssfam=\csname\vsSMALL bss\@nd
\def\bss{\fam\bssfam\csname\SMALL bss\@nd}%
\ifams
\textfont\msafam=\csname\SMALL msa\@nd
\scriptfont\msafam=\csname\smSMALL msa\@nd
\scriptscriptfont\msafam=\csname\vsSMALL msa\@nd
\textfont\msbfam=\csname\SMALL msb\@nd
\scriptfont\msbfam=\csname\smSMALL msb\@nd
\scriptscriptfont\msbfam=\csname\vsSMALL msb\@nd
\else\fi
\normalbaselineskip=\smallb@seline pt%
\setbox\strutbox=\hbox{\vrule height.7\normalbaselineskip
depth.3\baselineskip width0pt}%
\everymath{}%
\def\sc{\csname\smSMALL rm\@nd}\normalbaselines\rm}%
\everydisplay{\indenteddisplay
   \gdef\labeltype{\eqlabel}}%
%
%
\def\hexnumber@#1{\ifcase#1 0\or 1\or 2\or 3\or 4\or 5\or 6\or 7\or 8\or
 9\or A\or B\or C\or D\or E\or F\fi}
\edef\bffam@{\hexnumber@\bffam}
\edef\bdifam@{\hexnumber@\bdifam}
\edef\bsyfam@{\hexnumber@\bsyfam}
\def\undefine#1{\let#1\undefined}
\def\newsymbol#1#2#3#4#5{\let\next@\relax
 \ifnum#2=\thr@@\let\next@\bdifam@\else
 \ifams
 \ifnum#2=\@ne\let\next@\msafam@\else
 \ifnum#2=\tw@\let\next@\msbfam@\fi\fi
 \fi\fi
 \mathchardef#1="#3\next@#4#5}
\def\mathhexbox@#1#2#3{\relax
 \ifmmode\mathpalette{}{\m@th\mathchar"#1#2#3}%
 \else\leavevmode\hbox{$\m@th\mathchar"#1#2#3$}\fi}

\def\bi#1{{\fam\bdifam\relax#1}}
%
%
\ifams\input amsmacro\fi
%
%
\newsymbol\bitGamma 3000
\newsymbol\bitDelta 3001
\newsymbol\bitTheta 3002
\newsymbol\bitLambda 3003
\newsymbol\bitXi 3004
\newsymbol\bitPi 3005
\newsymbol\bitSigma 3006
\newsymbol\bitUpsilon 3007
\newsymbol\bitPhi 3008
\newsymbol\bitPsi 3009
\newsymbol\bitOmega 300A
\newsymbol\balpha 300B
\newsymbol\bbeta 300C
\newsymbol\bgamma 300D
\newsymbol\bdelta 300E
\newsymbol\bepsilon 300F
\newsymbol\bzeta 3010
\newsymbol\bfeta 3011
\newsymbol\btheta 3012
\newsymbol\biota 3013
\newsymbol\bkappa 3014
\newsymbol\blambda 3015
\newsymbol\bmu 3016
\newsymbol\bnu 3017
\newsymbol\bxi 3018
\newsymbol\bpi 3019
\newsymbol\brho 301A
\newsymbol\bsigma 301B
\newsymbol\btau 301C
\newsymbol\bupsilon 301D
\newsymbol\bphi 301E
\newsymbol\bchi 301F
\newsymbol\bpsi 3020
\newsymbol\bomega 3021
\newsymbol\bvarepsilon 3022
\newsymbol\bvartheta 3023
\newsymbol\bvaromega 3024
\newsymbol\bvarrho 3025
\newsymbol\bvarzeta 3026
\newsymbol\bvarphi 3027
\newsymbol\bpartial 3040
\newsymbol\bell 3060
\newsymbol\bimath 307B
\newsymbol\bjmath 307C
\mathchardef\binfty "0\bsyfam@31
\mathchardef\bnabla "0\bsyfam@72
\mathchardef\bdot "2\bsyfam@01
\mathchardef\bGamma "0\bffam@00
\mathchardef\bDelta "0\bffam@01
\mathchardef\bTheta "0\bffam@02
\mathchardef\bLambda "0\bffam@03
\mathchardef\bXi "0\bffam@04
\mathchardef\bPi "0\bffam@05
\mathchardef\bSigma "0\bffam@06
\mathchardef\bUpsilon "0\bffam@07
\mathchardef\bPhi "0\bffam@08
\mathchardef\bPsi "0\bffam@09
\mathchardef\bOmega "0\bffam@0A
\mathchardef\itGamma "0100
\mathchardef\itDelta "0101
\mathchardef\itTheta "0102
\mathchardef\itLambda "0103
\mathchardef\itXi "0104
\mathchardef\itPi "0105
\mathchardef\itSigma "0106
\mathchardef\itUpsilon "0107
\mathchardef\itPhi "0108
\mathchardef\itPsi "0109
\mathchardef\itOmega "010A
\mathchardef\Gamma "0000
\mathchardef\Delta "0001
\mathchardef\Theta "0002
\mathchardef\Lambda "0003
\mathchardef\Xi "0004
\mathchardef\Pi "0005
\mathchardef\Sigma "0006
\mathchardef\Upsilon "0007
\mathchardef\Phi "0008
\mathchardef\Psi "0009
\mathchardef\Omega "000A
%
%
\newcount\firstpage  \firstpage=1  
\newcount\jnl                      
\newcount\secno                    
\newcount\subno                    
\newcount\subsubno                 
\newcount\appno                    
\newcount\tabno                    
\newcount\figno                    
\newcount\countno                  
\newcount\refno                    
\newcount\eqlett     \eqlett=97    
\newif\ifletter
\newif\ifwide
\newif\ifnotfull
\newif\ifaligned
\newif\ifnumbysec
\newif\ifappendix
\newif\ifnumapp
\newif\ifssf
\newif\ifppt
\newdimen\t@bwidth
\newdimen\c@pwidth
\newdimen\digitwidth                    
\newdimen\argwidth                      
\newdimen\secindent    \secindent=5pc   
\newdimen\textind    \textind=16pt      
\newdimen\tempval                       
\newskip\beforesecskip
\def\beforesecspace{\vskip\beforesecskip\relax}
\newskip\beforesubskip
\def\beforesubspace{\vskip\beforesubskip\relax}
\newskip\beforesubsubskip
\def\beforesubsubspace{\vskip\beforesubsubskip\relax}
\newskip\secskip
\def\secspace{\vskip\secskip\relax}
\newskip\subskip
\def\subspace{\vskip\subskip\relax}
\newskip\insertskip
\def\insertspace{\vskip\insertskip\relax}
\def\sp@ce{\ifx\next*\let\next=\@ssf
               \else\let\next=\@nossf\fi\next}
\def\@ssf#1{\nobreak\secspace\global\ssftrue\nobreak}
\def\@nossf{\nobreak\secspace\nobreak\noindent\ignorespaces}
\def\subsp@ce{\ifx\next*\let\next=\@sssf
               \else\let\next=\@nosssf\fi\next}
\def\@sssf#1{\nobreak\subspace\global\ssftrue\nobreak}
\def\@nosssf{\nobreak\subspace\nobreak\noindent\ignorespaces}
\beforesecskip=24pt plus12pt minus8pt
\beforesubskip=12pt plus6pt minus4pt
\beforesubsubskip=12pt plus6pt minus4pt
\secskip=12pt plus 2pt minus 2pt
\subskip=6pt plus3pt minus2pt
\insertskip=18pt plus6pt minus6pt%
\fontdimen16\tensy=2.7pt
\fontdimen17\tensy=2.7pt
%
%
\def\eqlabel{(\ifappendix\applett
               \ifnumbysec\ifnum\secno>0 \the\secno\fi.\fi
               \else\ifnumbysec\the\secno.\fi\fi\the\countno)}
\def\seclabel{\ifappendix\ifnumapp\else\applett\fi
    \ifnum\secno>0 \the\secno
    \ifnumbysec\ifnum\subno>0.\the\subno\fi\fi\fi
    \else\the\secno\fi\ifnum\subno>0.\the\subno
         \ifnum\subsubno>0.\the\subsubno\fi\fi}
\def\tablabel{\ifappendix\applett\fi\the\tabno}
\def\figlabel{\ifappendix\applett\fi\the\figno}
\def\gac{\global\advance\countno by 1}
%
%

\def\vfootnote#1{\insert\footins\bgroup
\interlinepenalty=\interfootnotelinepenalty
\splittopskip=\ht\strutbox 
\splitmaxdepth=\dp\strutbox \floatingpenalty=20000
\leftskip=0pt \rightskip=0pt \spaceskip=0pt \xspaceskip=0pt%
\noindent\smallfonts\rm #1\ \ignorespaces\footstrut\futurelet\next\fo@t}
%
%
\def\endinsert{\egroup
    \if@mid \dimen@=\ht0 \advance\dimen@ by\dp0
       \advance\dimen@ by12\p@ \advance\dimen@ by\pagetotal
       \ifdim\dimen@>\pagegoal \@midfalse\p@gefalse\fi\fi
    \if@mid \insertspace \box0 \par \ifdim\lastskip<\insertskip
    \removelastskip \penalty-200 \insertspace \fi
    \else\insert\topins{\penalty100
       \splittopskip=0pt \splitmaxdepth=\maxdimen
       \floatingpenalty=0
       \ifp@ge \dimen@=\dp0
       \vbox to\vsize{\unvbox0 \kern-\dimen@}%
       \else\box0\nobreak\insertspace\fi}\fi\endgroup}
%
%
%
\def\ind{\hbox to \secindent{\hfill}}
%
%

%
%
\def\lo#1{\llap{${}#1{}$}}
%
%
\def\indeqn#1{\alignedfalse\displ@y\halign{\hbox to \displaywidth
    {$\ind\@lign\displaystyle##\hfil$}\crcr #1\crcr}}
%
%
\def\indalign#1{\alignedtrue\displ@y \tabskip=0pt
  \halign to\displaywidth{\ind$\@lign\displaystyle{##}$\tabskip=0pt
    &$\@lign\displaystyle{{}##}$\hfill\tabskip=\centering
    &\llap{$\@lign\hbox{\rm##}$}\tabskip=0pt\crcr
    #1\crcr}}
\def\fl{{\hskip-\secindent}}
\def\indenteddisplay#1$${\indispl@y{#1 }}
\def\indispl@y#1{\disptest#1\eqalignno\eqalignno\disptest}
\def\disptest#1\eqalignno#2\eqalignno#3\disptest{%
    \ifx#3\eqalignno
    \indalign#2%
    \else\indeqn{#1}\fi$$}
%
%

%
%

%
%

%
%

%
%

\def\ns{\noalign{\vskip-3pt}}

%

%
%
\def\bhbar{\rlap{\kern1pt\raise.4ex\hbox{\bf\char'40}}\bi{h}}

\def\etal{{\it et al\/}\ }
\def\frac#1#2{{#1\over#2}}
\ifams
\def\lap{\lesssim}
\def\gap{\gtrsim}

\else

\def\gap{\;\lower3pt\hbox{$\buildrel > \over \sim$}\;}%
\def\lap{\;\lower3pt\hbox{$\buildrel < \over \sim$}\;}\fi

\chardef\ii="10
\def\tqs{\hbox to 25pt{\hfil}}

\def\Bbbone{1\kern-.22em {\rm l}}
%
%
\def\rp{\raise8pt\hbox{$\scriptstyle\prime$}}
%
%
%
%

%
%
\def\[#1\]{\setbox0=\hbox{$\dsty#1$}\argwidth=\wd0
    \setbox0=\hbox{$\left[\box0\right]$}\advance\argwidth by -\wd0
    \left[\kern.3\argwidth\box0\kern.3\argwidth\right]}
%
%
\def\lsb#1\rsb{\setbox0=\hbox{$#1$}\argwidth=\wd0
    \setbox0=\hbox{$\left[\box0\right]$}\advance\argwidth by -\wd0
    \left[\kern.3\argwidth\box0\kern.3\argwidth\right]}
%

%
%

%
\def\pt(#1){({\it #1\/})}
\let\dsty=\displaystyle

%
%
\def\reactions#1{\vskip 12pt plus2pt minus2pt%
\vbox{\hbox{\kern\secindent\vrule\kern12pt%
\vbox{\kern0.5pt\vbox{\hsize=24pc\parindent=0pt\smallfonts\rm NUCLEAR
REACTIONS\strut\quad #1\strut}\kern0.5pt}\kern12pt\vrule}}}
%
%
\def\slashchar#1{\setbox0=\hbox{$#1$}\dimen0=\wd0%
\setbox1=\hbox{/}\dimen1=\wd1%
\ifdim\dimen0>\dimen1%
\rlap{\hbox to \dimen0{\hfil/\hfil}}#1\else
\rlap{\hbox to \dimen1{\hfil$#1$\hfil}}/\fi}
%
%
\def\textindent#1{\noindent\hbox to \parindent{#1\hss}\ignorespaces}
%
%
\def\opencirc{\raise1pt\hbox{$\scriptstyle{\bigcirc}$}}

\ifams
\def\opensqr{\hbox{$\square$}}

\def\opentridown{\hbox{$\triangledown$}}

\else
\def\opensqr{\vbox{\hrule height.4pt\hbox{\vrule width.4pt height3.5pt
    \kern3.5pt\vrule width.4pt}\hrule height.4pt}}

\def\opentridown{\raise1pt\hbox{$\scriptstyle\bigtriangledown$}}

\fi

%
%
\def\m@th{\mathsurround=0pt}
%
%
\def\cases#1{%
\left\{\,\vcenter{\normalbaselines\openup1\jot\m@th%
     \ialign{$\displaystyle##\hfil$&\rm\tqs##\hfil\crcr#1\crcr}}\right.}%
%
%
\def\oldcases#1{\left\{\,\vcenter{\normalbaselines\m@th
    \ialign{$##\hfil$&\rm\quad##\hfil\crcr#1\crcr}}\right.}
%
%
\def\numcases#1{\left\{\,\vcenter{\baselineskip=15pt\m@th%
     \ialign{$\displaystyle##\hfil$&\rm\tqs##\hfil
     \crcr#1\crcr}}\right.\hfill
     \vcenter{\baselineskip=15pt\m@th%
     \ialign{\rlap{$\phantom{\displaystyle##\hfil}$}\tabskip=0pt&\en
     \rlap{\phantom{##\hfil}}\crcr#1\crcr}}}
\def\ptnumcases#1{\left\{\,\vcenter{\baselineskip=15pt\m@th%
     \ialign{$\displaystyle##\hfil$&\rm\tqs##\hfil
     \crcr#1\crcr}}\right.\hfill
     \vcenter{\baselineskip=15pt\m@th%
     \ialign{\rlap{$\phantom{\displaystyle##\hfil}$}\tabskip=0pt&\enpt
     \rlap{\phantom{##\hfil}}\crcr#1\crcr}}\global\eqlett=97
     \global\advance\countno by 1}
%
%
\def\eq(#1){\ifaligned\@mp(#1)\else\hfill\llap{{\rm (#1)}}\fi}
\def\ceq(#1){\ns\ns\ifaligned\@mp\fi\eq(#1)\cr\ns\ns}
\def\eqpt(#1#2){\ifaligned\@mp(#1{\it #2\/})
                    \else\hfill\llap{{\rm (#1{\it #2\/})}}\fi}
\let\eqno=\eq
%
%
\countno=1

\def\aleq{&\rm(\ifappendix\applett
               \ifnumbysec\ifnum\secno>0 \the\secno\fi.\fi
               \else\ifnumbysec\the\secno.\fi\fi\the\countno}
\def\noaleq{\hfill\llap\bgroup\rm(\ifappendix\applett
               \ifnumbysec\ifnum\secno>0 \the\secno\fi.\fi
               \else\ifnumbysec\the\secno.\fi\fi\the\countno}
\def\@mp{&}
\def\en{\ifaligned\aleq)\else\noaleq)\egroup\fi\gac}
\def\cen{\ns\ns\ifaligned\@mp\fi\en\cr\ns\ns}
\def\enpt{\ifaligned\aleq{\it\char\the\eqlett})\else
    \noaleq{\it\char\the\eqlett})\egroup\fi
    \global\advance\eqlett by 1}
\def\endpt{\ifaligned\aleq{\it\char\the\eqlett})\else
    \noaleq{\it\char\the\eqlett})\egroup\fi
    \global\eqlett=97\gac}
%
%




%
%

\def\NP{{\it Nucl. Phys.}}
\def\PL{{\it Phys. Lett.}}

\def\RMP{{\it Rev. Mod. Phys.}}

\headline={\ifodd\pageno{\ifnum\pageno=\firstpage\hfill
   \else\rrhead\fi}\else\lrhead\fi}
\def\rrhead{\textfonts\hskip\secindent\it
    \shorttitle\hfill\rm\folio}
\def\lrhead{\textfonts\hbox to\secindent{\rm\folio\hss}%
    \it\aunames\hss}
\footline={\ifnum\pageno=\firstpage \hfill\textfonts\rm\folio\fi}
\def\@rticle#1#2{\vglue.5pc
    {\parindent=\secindent \bf #1\par}
     \vskip2.5pc
    {\exhyphenpenalty=10000\hyphenpenalty=10000
     \baselineskip=18pt\raggedright\noindent
     \headfonts\bf#2\par}\futurelet\next\sh@rttitle}%
\def\title#1{\gdef\shorttitle{#1}
    \vglue4pc{\exhyphenpenalty=10000\hyphenpenalty=10000
    \baselineskip=18pt
    \raggedright\parindent=0pt
    \headfonts\bf#1\par}\futurelet\next\sh@rttitle}

\def\article#1#2{\gdef\shorttitle{#2}\@rticle{#1}{#2}}
\def\review#1{\gdef\shorttitle{#1}%
    \@rticle{REVIEW \ifpbm\else ARTICLE\fi}{#1}}
\def\topical#1{\gdef\shorttitle{#1}%
    \@rticle{TOPICAL REVIEW}{#1}}
\def\comment#1{\gdef\shorttitle{#1}%
    \@rticle{COMMENT}{#1}}
\def\note#1{\gdef\shorttitle{#1}%
    \@rticle{NOTE}{#1}}
\def\prelim#1{\gdef\shorttitle{#1}%
    \@rticle{PRELIMINARY COMMUNICATION}{#1}}
\def\letter#1{\gdef\shorttitle{Letter to the Editor}%
     \gdef\aunames{Letter to the Editor}
     \global\lettertrue\ifnum\jnl=7\global\letterfalse\fi
     \@rticle{LETTER TO THE EDITOR}{#1}}
\def\sh@rttitle{\ifx\next[\let\next=\sh@rt
                \else\let\next=\f@ll\fi\next}
\def\sh@rt[#1]{\gdef\shorttitle{#1}}
\def\f@ll{}
\def\author#1{\ifletter\else\gdef\aunames{#1}\fi\vskip1.5pc
    {\parindent=\secindent
     \hang\textfonts
     \ifppt\bf\else\rm\fi#1\par}
     \ifppt\bigskip\else\smallskip\fi
     \futurelet\next\@unames}
\def\@unames{\ifx\next[\let\next=\short@uthor
                 \else\let\next=\@uthor\fi\next}
\def\short@uthor[#1]{\gdef\aunames{#1}}
\def\@uthor{}
\def\address#1{{\parindent=\secindent
    \exhyphenpenalty=10000\hyphenpenalty=10000
\ifppt\textfonts\else\smallfonts\fi\hang\raggedright\rm#1\par}%
    \ifppt\bigskip\fi}
\def\jl#1{\global\jnl=#1}
\jl{0}%
\def\journal{\ifnum\jnl=1 J. Phys.\ A: Math.\ Gen.\
        \else\ifnum\jnl=2 J. Phys.\ B: At.\ Mol.\ Opt.\ Phys.\
        \else\ifnum\jnl=3 J. Phys.:\ Condens. Matter\
        \else\ifnum\jnl=4 J. Phys.\ G: Nucl.\ Part.\ Phys.\
        \else\ifnum\jnl=5 Inverse Problems\
        \else\ifnum\jnl=6 Class. Quantum Grav.\
        \else\ifnum\jnl=7 Network\
        \else\ifnum\jnl=8 Nonlinearity\
        \else\ifnum\jnl=9 Quantum Opt.\
        \else\ifnum\jnl=10 Waves in Random Media\
        \else\ifnum\jnl=11 Pure Appl. Opt.\
        \else\ifnum\jnl=12 Phys. Med. Biol.\
        \else\ifnum\jnl=13 Modelling Simulation Mater.\ Sci.\ Eng.\
        \else\ifnum\jnl=14 Plasma Phys. Control. Fusion\
        \else\ifnum\jnl=15 Physiol. Meas.\
        \else\ifnum\jnl=16 Sov.\ Lightwave Commun.\
        \else\ifnum\jnl=17 J. Phys.\ D: Appl.\ Phys.\
        \else\ifnum\jnl=18 Supercond.\ Sci.\ Technol.\
        \else\ifnum\jnl=19 Semicond.\ Sci.\ Technol.\
        \else\ifnum\jnl=20 Nanotechnology\
        \else\ifnum\jnl=21 Meas.\ Sci.\ Technol.\
        \else\ifnum\jnl=22 Plasma Sources Sci.\ Technol.\
        \else\ifnum\jnl=23 Smart Mater.\ Struct.\
        \else\ifnum\jnl=24 J.\ Micromech.\ Microeng.\
   \else Institute of Physics Publishing\
   \fi\fi\fi\fi\fi\fi\fi\fi\fi\fi\fi\fi\fi\fi\fi
   \fi\fi\fi\fi\fi\fi\fi\fi\fi}
\def\beginabstract{\insertspace
     \parindent=\secindent\ifppt\textfonts\else\smallfonts\fi
     \hang{\bf Abstract. }\rm }
\def\endabstract{\par
    \parindent=\textind\textfonts\rm
    \ifppt\vfill\fi}

\def\submitted{\ifppt\noindent\textfonts\rm Submitted to \journal\par
     \bigskip\fi}
\def\today{\number\day\ \ifcase\month\or
     January\or February\or March\or April\or May\or June\or
     July\or August\or September\or October\or November\or
     December\fi\space \number\year}
\def\date{\ifppt\noindent\textfonts\rm
     Date: \today\par\goodbreak\bigskip\fi}
%
%

%

%
%
\def\section#1{\ifppt\ifnum\secno=0\eject\fi\fi
    \subno=0\subsubno=0\global\advance\secno by 1
    \gdef\labeltype{\seclabel}\ifnumbysec\countno=1\fi
    \goodbreak\beforesecspace\nobreak
    \noindent{\bf \the\secno. #1}\par\futurelet\next\sp@ce}
\def\subsection#1{\subsubno=0\global\advance\subno by 1
     \gdef\labeltype{\seclabel}%
     \ifssf\else\goodbreak\beforesubspace\fi
     \global\ssffalse\nobreak
     \noindent{\it \the\secno.\the\subno. #1\par}%
     \futurelet\next\subsp@ce}
\def\subsubsection#1{\global\advance\subsubno by 1
     \gdef\labeltype{\seclabel}%
     \ifssf\else\goodbreak\beforesubsubspace\fi
     \global\ssffalse\nobreak
     \noindent{\it \the\secno.\the\subno.\the\subsubno. #1}\null.
     \ignorespaces}
\def\nosections{\ifppt\eject\else\vskip30pt plus12pt minus12pt\fi
    \noindent\ignorespaces}
%
%
\def\numappendix#1{\ifappendix\ifnumbysec\countno=1\fi\else
    \countno=1\figno=0\tabno=0\fi
    \subno=0\global\advance\appno by 1
    \secno=\appno\gdef\applett{A}\gdef\labeltype{\seclabel}%
    \global\appendixtrue\global\numapptrue
    \goodbreak\beforesecspace\nobreak
    \noindent{\bf Appendix \the\appno. #1\par}%
    \futurelet\next\sp@ce}
\def\numsubappendix#1{\global\advance\subno by 1\subsubno=0
    \gdef\labeltype{\seclabel}%
    \ifssf\else\goodbreak\beforesubspace\fi
    \global\ssffalse\nobreak
    \noindent{\it A\the\appno.\the\subno. #1\par}%
    \futurelet\next\subsp@ce}
\def\@ppendix#1#2#3{\countno=1\subno=0\subsubno=0\secno=0\figno=0\tabno=0
    \gdef\applett{#1}\gdef\labeltype{\seclabel}\global\appendixtrue
    \goodbreak\beforesecspace\nobreak
    \noindent{\bf Appendix#2#3\par}\futurelet\next\sp@ce}
\def\Appendix#1{\@ppendix{A}{. }{#1}}
\def\appendix#1#2{\@ppendix{#1}{ #1. }{#2}}
\def\App#1{\@ppendix{A}{ }{#1}}
\def\app{\@ppendix{A}{}{}}
\def\subappendix#1#2{\global\advance\subno by 1\subsubno=0
    \gdef\labeltype{\seclabel}%
    \ifssf\else\goodbreak\beforesubspace\fi
    \global\ssffalse\nobreak
    \noindent{\it #1\the\subno. #2\par}%
    \nobreak\subspace\noindent\ignorespaces}
%
%
\def\@ck#1{\ifletter\bigskip\noindent\ignorespaces\else
    \goodbreak\beforesecspace\nobreak
    \noindent{\bf Acknowledgment#1\par}%
    \nobreak\secspace\noindent\ignorespaces\fi}
\def\ack{\@ck{s}}
\def\ackn{\@ck{}}
\def\n@ip#1{\goodbreak\beforesecspace\nobreak
    \noindent\smallfonts{\it #1}. \rm\ignorespaces}
\def\naip{\n@ip{Note added in proof}}
\def\na{\n@ip{Note added}}

%
%

%

%
%

%

%

\def\tablecont{\topinsert\global\advance\tabno by -1
    \tablecaption{(continued)}}
\def\tablecaption#1{\gdef\labeltype{\tablabel}\global\widefalse
    \leftskip=\secindent\parindent=0pt
    \global\advance\tabno by 1
    \smallfonts{\bf Table \ifappendix\applett\fi\the\tabno.} \rm #1\par
    \smallskip\futurelet\next\t@b}
\def\t@b{\ifx\next*\let\next=\widet@b
             \else\ifx\next[\let\next=\fullwidet@b
                      \else\let\next=\narrowt@b\fi\fi
             \next}
\def\widet@b#1{\global\widetrue\global\notfulltrue
    \t@bwidth=\hsize\advance\t@bwidth by -\secindent}
\def\fullwidet@b[#1]{\global\widetrue\global\notfullfalse
    \leftskip=0pt\t@bwidth=\hsize}
\def\narrowt@b{\global\notfulltrue}
\def\align{\catcode`?=13\ifnotfull\moveright\secindent\fi
    \vbox\bgroup\halign\ifwide to \t@bwidth\fi
    \bgroup\strut\tabskip=1.2pc plus1pc minus.5pc}
\def\endalign{\egroup\egroup\catcode`?=12}

%
%

%
%

%

%
%

%

\catcode`?=13
\def\lineup{\setbox0=\hbox{\smallfonts\rm 0}%
    \digitwidth=\wd0%
    \def?{\kern\digitwidth}%
    \def\\{\hbox{$\phantom{-}$}}%
    \def\-{\llap{$-$}}}
\catcode`?=12
%
%
\def\sidetable#1#2{\hbox{\ifppt\hsize=18pc\t@bwidth=18pc
                          \else\hsize=15pc\t@bwidth=15pc\fi
    \parindent=0pt\vtop{\null #1\par}%
    \ifppt\hskip1.2pc\else\hskip1pc\fi
    \vtop{\null #2\par}}}
\def\lstable#1#2{\everypar{}\tempval=\hsize\hsize=\vsize
    \vsize=\tempval\hoffset=-3pc
    \global\tabno=#1\gdef\labeltype{\tablabel}%
    \noindent\smallfonts{\bf Table \ifappendix\applett\fi
    \the\tabno.} \rm #2\par
    \smallskip\futurelet\next\t@b}
\def\inctabno{\global\advance\tabno by 1}
%
%

%

%
\def\figure#1{\figc@ption{#1}\bigskip}
\def\figc@ption#1{\global\advance\figno by 1\gdef\labeltype{\figlabel}%
   {\parindent=\secindent\smallfonts\hang
    {\bf Figure \ifappendix\applett\fi\the\figno.} \rm #1\par}}
%
%
\def\refHEAD{\goodbreak\beforesecspace
     \noindent\textfonts{\bf References}\par
     \let\ref=\rf
     \nobreak\smallfonts\rm}
\def\references{\refHEAD\parindent=0pt
     \everypar{\hangindent=18pt\hangafter=1
     \frenchspacing\rm}%
     \secspace}
\def\rf#1{\par\noindent\hbox to 21pt{\hss #1\quad}\ignorespaces}
\def\refjl#1#2#3#4{\noindent #1 {\it #2 \bf #3} #4\par}
\def\refbk#1#2#3{\noindent #1 {\it #2} #3\par}
%
%

%
%

%
%

\def\Proof{\bigbreak\noindent{\it Proof}.\quad\ignorespaces}

%
%

%
\catcode`\@=12
%
%
\def\pptstyle{\ppttrue\headsize{17}{24}%
\textsize{12}{16}%
\smallsize{10}{12}%
\hsize=37.2pc\vsize=56pc
\textind=20pt\secindent=6pc}
%
%

%
%

%
%

%
\parindent=\textind
%
%

\def\Proposition{\bigbreak\noindent{\it Proposition}.\quad\ignorespaces}

\pptstyle
\jl{6} 


\letter{Normal frames for non-Riemannian connections}

\author{David Hartley\footnote{\dag}{Email: hartley@gmd.de}
\footnote{\ddag}{Postal address: GMD--SCAI, D--53754 St~Augustin, Germany}}

\address{Institute for Algorithms and Scientific Computing,
GMD --- German National Research Center for Information Technology,
St~Augustin, Germany}

\beginabstract
The principal properties of geodesic normal coordinates are the vanishing
of the connection components and first derivatives of the metric components
at some point. It is well-known that these hold only at points where the
connection has vanishing torsion and non-metricity.  However, it is shown
that normal frames, possessing the essential features of normal
coordinates, can still be constructed when the connection is
non-Riemannian.
\endabstract

\submitted
\date


\nosections


Almost the defining property of general relativity is its coordinate and
frame independence, a feature which becomes obvious when tensor notation is
used.  Nevertheless, it is frequently convenient to single out certain
coordinate and frame choices in order to simplify particular calculations
or arguments, or to make approximations to compare with experiment. In many
places, for example in comparisons with special relativity or arguments
motivated by the equivalence principle, an inertial frame of reference is
desirable.

Unfortunately, curved space-time does not admit inertial frames of
reference in general, so geodesic normal coordinates are commonly used to
provide an instantaneous inertial frame at a single space-time point.  In
these coordinates, the metric components are stationary and the connection
components vanish at the chosen point. As is common knowledge, these
features are lost when the connection is non-Riemannian. For example, it is
clear from the definition of the torsion operator $\tau$ on vector fields
$X$, $Y$,

$$	\tau(X,Y) = \nabla_X Y - \nabla_Y X - [X,Y]		\eqno(1)
$$

that if $X$ and $Y$ are coordinate vectors for which the connection
components vanish at some point, then the torsion must vanish at the point
too. Likewise, the non-metricity

$$
\fl	Q(X,Y,Z) = \nabla_{\lower.5ex\hbox{$\scriptstyle X$}} g(Y,Z)\cr
	      \lo= X\big(g(Y,Z)\big) - g(\nabla_XY,Z) - g(Y,\nabla_XZ)
								\eqno(2)
$$

reduces to first derivatives of the metric components in coordinates where
the connection components vanish.

It is, of course, still possible to construct normal coordinates using
the exponential map based on autoparallels of a non-Riemannian
connection. However, for non-vanishing torsion, the connection components
will not vanish, and in the presence of non-metricity, the metric
components will not be stationary.

The purpose of this note is to point out that, notwithstanding the failure
of normal coordinates, it is still possible to define a normal frame at a
point for a non-Riemannian connection. Such a possibility has been
mentioned before within the context of the equivalence principle for
theories with torsion (von der Heyde 1975, Hehl \etal 1976), but without a
full proof. Related concepts for gauge theories (using Riemann normal
coordinates) have also been employed in heat-kernel studies on
Riemann-Cartan space-time (Obukhov 1983, Cognola and Zerbini 1988).



Let $M$ be a differentiable manifold, and $\nabla$ an arbitrary connection
on $M$. Take an arbitrary point $p\in M$, and let $\{X_i\}_p$ be a basis
for $T_pM$.

\Proposition
$\{X_i\}_p$ can be extended to a frame $\{X_i\}$ on some neighbourhood
$U\subset M$ of $p$ in such a way that $\nabla X_i = 0$ at $p$ for each $i$.

\Proof
For each $i$, there exists an autoparallel $C_i$ in a simply connected
neighbourhood $U_i$ of $p$ which passes through $p$ with tangent vector
$X_i$ at $p$. For each $i$, define a set of vector fields $\{X_j\}$ along
$C_i$ by parallel transporting $\{X_j\}_p$. By suitably restricting the
intersection $U$ of the $U_i$, the autoparallels will not meet, and the
vector fields can be extended arbitrarily to form a frame on $U$. Since, by
construction, $\nabla_{X_i}X_j = 0$ at $p$ for all $i,j$, it follows that
$\nabla X_i = 0$ at $p$.

This result makes no mention of any metric on $M$, it is simply a property
of connections. Now let $g$ be a metric on $M$. A frame $\{X_i\}$ is said
to be {\it normal} (Sachs and Wu 1977) at $p$ if $\{X_i\}_p$ is orthonormal
and $\nabla X_i = 0$ at $p$. Starting with an orthonormal basis at any
point $p$ in $M$, it is always possible to extend it using the above method
to a local normal frame with respect to any connection. If the connection
and metric are compatible ($\nabla g=0$), the normal frame can be made
orthonormal away from $p$ as well.

This result can be expressed using the connection components defined by
$\nabla_{X_i} X_j = \Gamma^k{}_{ij} X_k$. Furthermore, it is obvious that
there always exist coordinates $\{x^i\}$ covering $p$ such that
$\partial_{x^i} = X_i$ at $p$. The component formulation is then

\Proposition
Let $M$ be a manifold with an arbitrary connection. For any point $p\in M$,
there exist coordinates $\{x^i\}$ and a frame $\{X_i\}$ in a neighbourhood
of $p$ such that, at $p$,
$$
	X_i = \partial_{x^i}\cr					\ceq(3)
	\Gamma^k{}_{ij} = 0,\cr
$$
where $\Gamma^k{}_{ij}$ are the connection components referred to the frame
$\{X_i\}$.

Note that the connection components referred to the frame
$\{\partial_{x^i}\}$ need not vanish (even at $p$). They will necessarily
be non-zero if the torsion is non-vanishing at $p$.


The essential properties of normal coordinates are the vanishing of the
connection components and first derivatives of the metric components. Both
of these are properties in the tangent spaces, not on the manifold
itself. So a normal frame suffices for most discussions where normal
coordinates are used.  Since normal frames may be defined for
non-Riemannian connections also, these same discussions can be adapted to
cover non-Riemannian gravity as well.

This will be most successful in Riemann-Cartan space-time, where torsion is
present but the connection is metric compatible. Here, the normal frame may
also be made orthonormal. This is desirable if, for instance, the
discussion involves spinors.

It is interesting to note that, as follows easily from (1), the existence
of an anholonomic normal frame implies the presence of torsion. Likewise,
{}from (2), a normal frame for which the metric components are not stationary
requires non-metricity. Turned around, this means that the only normal
frames for a Riemannian connection are those generated by normal coordinates.

In this note, the existence of normal frames for space-times with
non-Riemannian connections has been proven by construction. These frames
have been argued to retain the salient features of normal coordinates in
general relativity, and thus allow similar simplifications and
approximations to be made in non-Riemannian gravity.

\ack

It is a pleasure to thank F~W~Hehl, F~Gronwald and P~A~Tuckey for helpful
discussions.  The work described here was carried out with the support of
the Graduate College on Scientific Computing, University of Cologne and GMD
St Augustin, funded by the German Research Foundation (DFG).

\references

\refjl{von der Heyde P 1975}{\it Lett. Nuovo Cimento}{14}{250--252}
\refjl{Hehl F W, von der Heyde P, Kerlick G D and Nester J M
 1976}{\RMP}{48}{393--416}
\refjl{Obukhov Y N 1983}{\NP}{B212}{237--254}
\refjl{Cognola G and Zerbini S 1988}{\PL}{B214}{70--74}
\refbk{Sachs R K and Wu H 1977}{General Relativity for
Mathematicians}{(New York: Springer-Verlag) p. 79}

\bye